# High Field (14Tesla) Magneto Transport of Sm/PrFeAsO

R. S. Meena,[1,2] Shiva Kumar Singh,[1*] Anand Pal,[1] Anuj Kumar,[1] R. Jha,[1] K. V. R. Rao,[2] Yi. Du,[3] X. L. Wang[3] and V.P.S Awana[1]†

[1] Quantum Phenomena and Application, National Physical Laboratory (CSIR), New Delhi-110012, India

[2] Department of Physics, University of Rajasthan Jaipur, India

[3] Australian Institute of Innovative Materials, University of Wollongong, NSW 2522, Australia

We report high field magneto transport of Sm/PrFeAsO. Below spin density wave transition ($T_{SDW}$), the magneto-resistance (*MR*) of Sm/PrFeAsO is positive and increasing with decreasing temperature. The *MR* of SmFeAsO, is found 16%, whereas the same is 21.5% in case of PrFeAsO, at 2.5 K under applied magnetic field of 14 Tesla (T). In case of SmFeAsO, the variation of isothermal *MR* with field below 20 K is nonlinear at lower magnetic fields (< 2 Tesla) and the same is linear at moderately higher magnetic fields ($H \geq 3.5$ T). On the other hand PrFeAsO shows almost linear *MR* at all temperatures below 20 K. The anomalous behavior of *MR* being exhibited in PrFeAsO is originated from Dirac cone states. The stronger interplay of Fe and Pr ordered moments is responsible for this distinct behavior. PrFeAsO also shows a hump in resistivity (*R-T*) with possible conduction band (FeAs) mediated ordering of Pr moments at around 12 K. However the same is absent in SmFeAsO even down to 2 K. Our results of high field magneto-transport of up to 14 Tesla brings about clear distinction between ground states of SmFeAsO and PrFeAsO.

## 1. Introduction

Iron based superconducting pnictides have attracted a lot of interest due to the interplay of multi-band structure of Fermi surface and anti-ferromagnetism being mediated by the magnetic Fe ions [1]. To understand the framework that whether or not the Fe pnictides are strongly correlated systems like the cuprates, more need to be addressed. Since density-functional theory (*DFT*) calculations have indicated that the electron-phonon interaction is too weak to account for high transition temperatures [2-3], the strength of the Coulomb correlations could give some information related to the pairing mechanism in these compounds [4]. The undoped arsenides *RE*FeAsO (*RE* = La, Sm, Nd, Pr and Ce) and $A$Fe$_2$As$_2$ (A=Ba, Ca and Sr) are semimetal. Almost each *RE*FeAsO exhibits a structural phase transition followed by an anti-ferromagnetic spin-density wave (*SDW*) magnetic ordering ($T_{SDW}$) at around 150 K [5]. On the other- hand undoped chalcogenide (FeTe) shows anti-ferromagnetic *SDW* with $T_{SDW}$ at around 78 K [6]. It is known that the *MR* is a very powerful tool to investigate the electronic scattering and the topology of the Fermi surface. For example, in MgB$_2$, a large *MR* was found which is closely related to the multiband property [7]. The magnetoresistance can provide information about Dirac cone states. The Dirac cone state is a novel electronic state with ideal massless fermion character. It is theoretically predicted that Dirac cone states exist in iron pnictide superconductors via special band folding below the antiferromagnetic transition temperature [8-10] and is experimentally confirmed in BaFe$_2$As$_2$ [11-13]. Very high transport mobility leads to linear relationship between momentum and energy. This is due to the zero effective mass and the long relaxation time of the conduction electrons regardless of impurities and/or various many-body effects [14]. Landau level (*LL*) splittings of the Dirac cone states are proportional to the square root of the external magnetic field $H$ [$\Delta_n = \pm v_F(heH|n|/\pi)^{1/2}$ where $v_F$ is the Fermi velocity]. Behaviour of Dirac fermions under magnetic field is discussed in ref [15] and the energy scale associated with the Dirac fermions is rather different from the ordinary 2D electron gas. Thus energy scaling makes the *LL* states thermally stable even in moderate fields ($H \leq 10$ T) [15]. Consequently the low energy properties of discrete *LL* states become accessible to conventional experimental probes, especially in *MR* measurements.

It is reported that *MR* of PrFeAsO and BaFe$_2$As$_2$ is linear in lower temperature range and at low field [13, 16], whereas NdFeAsO doesn't show any linearity [17]. Abrikosov interpreted the phenomenon of linear *MR* by considering a quantum limit where all of the carriers in the Dirac cone states occupy only the zeroth *LL* [18]. This situation can be realized in two specific conditions: (1) When Fermi energy $E_F$ of the system is lower than the *LL* splitting $\Delta_1 = \pm v_F(heH/\pi)^{1/2}$ between the first and the zeroth *LL*s. This means that when *H* is higher than a critical value $H^*(0)$ at 0 K, all the carriers can occupy only the zeroth *LL*. (2) at the finite temperature, thermal fluctuation ($k_BT$) should not exceed $\Delta_1$. In such type of quantum limit, *MR* can no longer be described within the conventional framework of Born scattering approximation [13]. Instead *MR* is directly proportional to $(N_i/en^2)H$. In this situation the electron density *n* (Dirac carriers) and the scattering centers $N_i$ determine the *MR* [18]. Thus the resulting *MR* is linear in relation to *H*.

In pnictides of *1111* family the phenomena of linear MR is only observed in PrFeAsO [16]. This shows that the ground state of PrFeAsO is something different than other members of *1111* family [17]. In the present study we tried to investigate the cause behind it. It is interesting that Pr being non-magnetic (para-magnetic above 1 K), orders anti-ferromagnetically at 12 K in PrFeAsO. On the other-hand Sm is antiferromagnetic (*AFM*) below 15 K, but it orders anti-ferromagnetically at around 5 K in SmFeAsO. Earlier, muon-spin relaxation measurements had been made on REFeAsO (R = La, Ce, Pr, and Sm) compounds. In case of CeFeAsO considerable interaction between the RE and Fe magnetism below the ordering of Fe moments ($T = 140$ K) was found [19]. The resonant scattering experiments showed strong interplay between Fe and Sm magnetism in SmFeAsO [20]. The neutron diffraction studies showed a delicate interplay of Fe and Pr moments in PrFeAsO [21]. In case of NdFeAsO, a change from *AFM* to *FM* arrangement along the *c* direction below 15 K, accompanied with the onset of Nd *AFM* order below $T_{Nd}$ 6 K, is observed in neutron diffraction study [22]. Thus it is clear that the magnetism of REFeAsO is complex [19-22].

In the present article we studied the ground state magneto transport properties of arsenides (Sm/PrFeAsO) to unearth some of their complex magnetic peculiarities. Our results of high field (14 Tesla) magneto transport bring out clear distinction between ground states of superconducting arsenides (Sm/PrFeAsO).

2. **Experimental**

All the studied polycrystalline samples are synthesized through single step solid-state reaction route via vacuum encapsulation technique. Stoichiometric amounts of Fe, As, RE, and $Fe_2O_3$ are weighed and mixed. The mixed powders were palletized and vacuum-sealed ($10^{-4}$ Torr) in a quartz tube. These sealed quartz ampoules are placed in box furnace and heat treated at 550°C for 12 hours, 850°C for 12 hours and then at 1150°C for 33 hours in continuum. Finally furnace is allowed to cool down to room temperature at a rate of 1°C/minute. As synthesized samples are single phase in nature, with only minute impurity of $RE_2O_3$ (~ 4%), XRD patterns are not shown here. The resistivity measurements are carried out on Quantum design Physical Property Measurement System (PPMS-14T).

## 3. Results and discussion

Fig. 1 shows the resistivity versus temperature plots of synthesized SmFeAsO and PrFeAsO samples in zero field. The resistivity decreases slowly down to 200 K, indicating near metallic behavior. A broad turn is observed at around 150 K with a sharp metallic step, which is associated with both the structural phase transition from tetragonal to orthorhombic symmetry and the *SDW* magnetic transition ($T_N$) of Fe moments [1]. Another broad turn near 70 K is observed which may be due to impurity phase of $RE_2O_3$. The lower inset of Fig. 1 shows the possible antiferromagnetic ordering of Pr at around 12 K ($T_N^{Pr}$) in PrFeAsO, which is seen more clearly in its dρ/dT [see upper inset of Fig.1]. Interestingly, the same (resistivity step/anomaly) is absent in SmFeAsO down to lowest studied temperature of 2 K.

Figures 2a and Fig. 2b show the dependence of *MR*% with field for SmFeAsO and PrFeAsO respectively, at various temperatures below 200 K. *MR* is a very powerful tool to investigate the electronic scattering process and the information about the Fermi surface. *MR* is defined as

$$MR\ (H) = \Delta\rho/\rho(0)\ldots\ldots..(1)$$

Where, $\Delta\rho = \rho(H) - \rho(0)$, $\rho(H)$ is the resistivity in applied field $H$ and $\rho_0$ is the resistivity at zero field.

The change in *MR* from 200 K-150 K, is less than 2% even in the magnetic field of 14 Tesla for both the samples. With the structural transition and consequent anti-ferromagnetic *SDW* ordering below 150 K, the *MR*% increases rapidly and reaches 16% for SmFeAsO (Fig. 2a) and 21.5% for PrFeAsO (Fig. 2b) at 2.5 K under applied field of 14 Tesla. In SmFeAsO variation of *MR* with field is non-linear. Though the linearity of *MR* increases with decreasing temperature but at lower fields it remains non-linear even at 2.5 K. The observed variation of *MR* on *H* is linear in H of strength ($|H| \geq 3.5$ T), but changed from a linear to a quadratic relation (*MR* α $H^2$) in lower fields ($|H| \leq 1.0$ T) [Fig. 3a] at 2.5 K. Similar behaviour of *MR* is observed in BaFe$_2$As$_2$ but in different field range [13].

On the other-hand non-linearity of PrFeAsO disappears below 40 K and a linear variation of *MR* is reported with field [16]. We also found similar behavior in our PrFeAsO sample with maximum *MR* change of 21.5% at 5 K and 14 Tesla. The *MR* versus *H* curve develops a weak negligible curvature in the low-field region, which indicates a crossover to a quadratic behaviour as $H \rightarrow 0$. The dependence of *MR* on *H* is linear in field strength $|H| \geq 1.5$ T and changed from a linear to quadratic (*MR* α $H^2$) for lower *H* values ($|H| \leq 0.5$ T) [Fig. 3b] at 20 K. The *MR* at 5 K and 2.5 K increases linearly at both below and above 6 Tesla but with slightly different slopes [Fig. 3c]. This behavior signals a weak meta-magnetic transition [16]. A crossover from linear *H* dependence at high field to super-linear *H* dependence of *MR* at low fields is reported for PrFeAsO [16] and other materials [23]. The linear *MR* for PrFeAsO at lower temperatures can be explained as the inherent quantum limit of the zeroth Landau level (*LL*) of the Dirac cone states in accordance with Abrikosov's model of a quantum *MR* [13, 16, 18]. It is worthwhile to mention that impurities and/or various many-body effects don't have any impact on this behaviour [14]. Our results also support this fact as having almost equal amount of impurities, PrFeAsO shows quantum *MR* while the same is absent in SmFeAsO. Thus it is inherent property of particular compound. In the parent compound of pnictides, a Dirac surface state is created due to the *SDW* band reconstruction and the apex of the Dirac dispersion intersects the Fermi energy, giving rise to electron and hole pockets. In ref. [13] it is argued that, for PrFeAsO, the carrier density may be high enough to induce the existence of small pocket on the Fermi surface satisfying the quantum condition.

The *MR* behaviour of Sm/PrFeAsO below 40 K shows a clear distinction between ground states of different superconducting arsenides. Besides linear *MR*, slightly different slopes are also observed below and above 6 Tesla at 5 and 2.5 K in PrFeAsO [Fig. 3c]. The distinct behaviour of PrFeAsO can be interpreted with help of the neutron diffraction data below 40 K of ref. [21]. It is reported that the ordering of Fe moments along the *a* axis starts at around 85 K and reaches a maximum at ~ 40 K, below which an anomalous expansion of the *c* axis sets in. This expansion results in a negative thermal volume expansion of 0.015% at 2 K. It was proposed that this effect, which is absent in superconducting samples, is driven by a delicate interplay between Fe and Pr ordered moments [21]. The *RE* and Fe moments interplay is also observed in SmFeAsO and NdFeAsO [20, 22]. The interaction between two magnetic sublattices is observed at ~ 110 K for SmFeAsO with $T_N^{Sm}$ ~ 5K and for NdFeAsO ~ 15 K with $T_N^{Nd}$ ~ 6K [20-22]. However $T_N^{Pr}$ is 12 K in case of PrFeAsO. It seems interplay of Fe and Pr ordered moments is stronger than interplay of Fe and Sm/Nd. This interplay leads to a negative thermal volume expansion of PrFeAsO [21] and the same is absent in Sm/NdFeAsO. It is argued that some materials behave as a gapless semiconductor at a proper ''tuning'', e.g., by pressure or doping at a random point in the reciprocal space [18, 24]. This leads to a linear energy (momentum) spectrum with the contact of valence band to the conduction band, or slight hybridization of both, leading to the quantum magneto-resistance. Linear *MR* is observed below ~ 40 K with the occurrence of negative thermal volume expansion in PrFeAsO. The effect is obviously missing in case of SmFeAsO due to lack of negative thermal expansion.

### 4. Conclusion

High field magneto transport of parent compounds of arsenides (Sm/PrFeAsO) is studied. *MR* of SmFeAsO, is found to be as large as 16% whereas it reaches 21.5% in case of PrFeAsO, at 2.5 K and in magnetic field of 14 Tesla (T). Nonlinear variation of *MR* at low magnetic fields tends to be linear at moderate magnetic fields ($H \geq 3.5$ T) and at lower temperatures in SmFeAsO. In PrFeAsO the *MR* is linear in H of strength ($|H| \geq 1.5$ T) and changed from linear to a quadratic relation in lower fields ($|H| \leq 0.5$ T) at 20 K. At 5 K and 2.5 K, slight change in slopes of *MR* is observed with field (below and above 6 T). The distinct behaviour of *MR* in PrFeAsO than other arsenides (Sm/NdFeAsO) is due to stronger interplay of Fe and Pr ordered moments than that of interplay of Fe and Sm/Nd ordered moments. The results bring out clear distinction between ground state of superconducting arsenides.

**Acknowledgements**

The authors would like to thank DNPL Prof. R. C. Budhani for his constant support and encouragement. Authors S. K. Singh, A. Pal and A. Kumar would like to acknowledge *CSIR*, India, for providing fellowships.

**Figure Caption**

Fig. 1 Resistivity curve of the undoped Sm/PrFeAsO samples in zero magnetic field below 200 K. Lower inset shows resistivity of Sm/PrFeAsO from 30 K- 2 K the antiferromagnetic ordering of Pr at around 12 K ($T_N^{Pr}$) in PrFeAsO can be seen. Upper inset shows $d\rho/dT$ of PrFeAsO from 30 K- 2 K.

Fig. 2(a) Variation of *MR* with magnetic field (0-14T) at different temperatures in SmFeAsO. (b) Variation of *MR* with magnetic field (0-14T) of PrFeAsO at different temperatures.

Fig. 3(a) *dMR/dH* at 2.5 K and 100 K; at 2.5 K *MR* exhibits linear $H$ dependence for $H \geq 3.5$ T. (b) *dMR/dH* at 20 K and 100 K; at 20 K *MR* exhibits linear $H$ dependence for $H \geq 1.5$ T. (c) *MR* at 5 K and 2.5 K which shows linear increase with $H$ up to 6 T and it follows a steeper slope above this field.

Fig.1

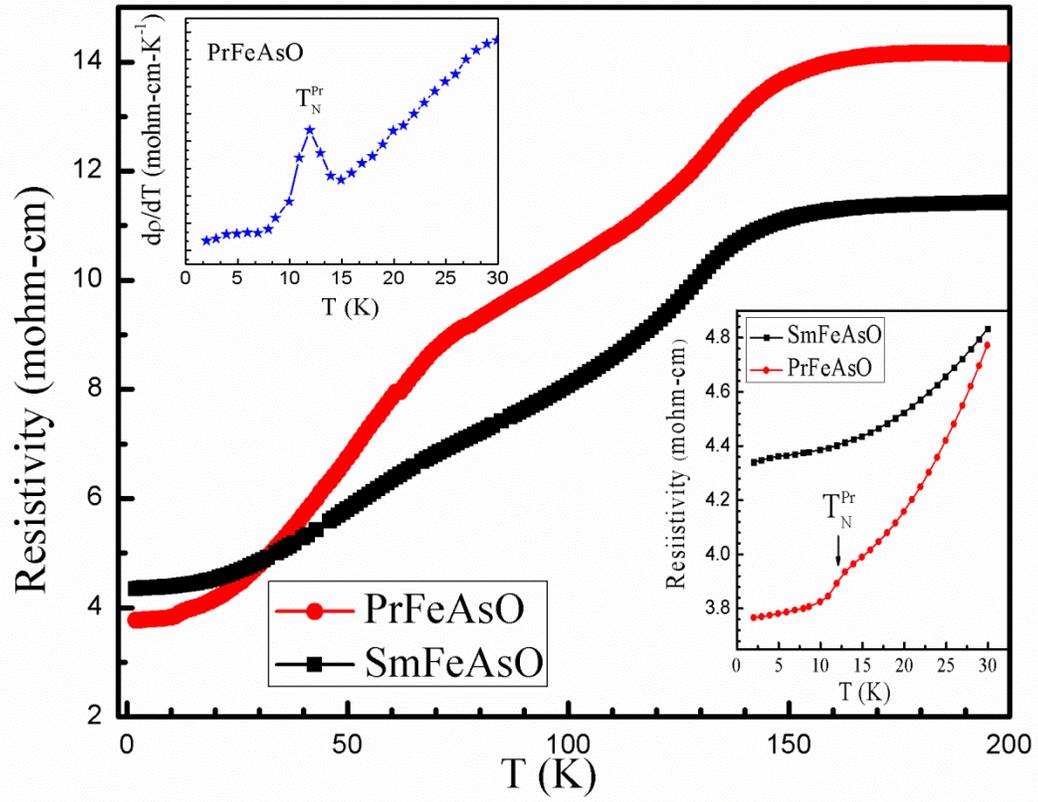

Fig. 2a

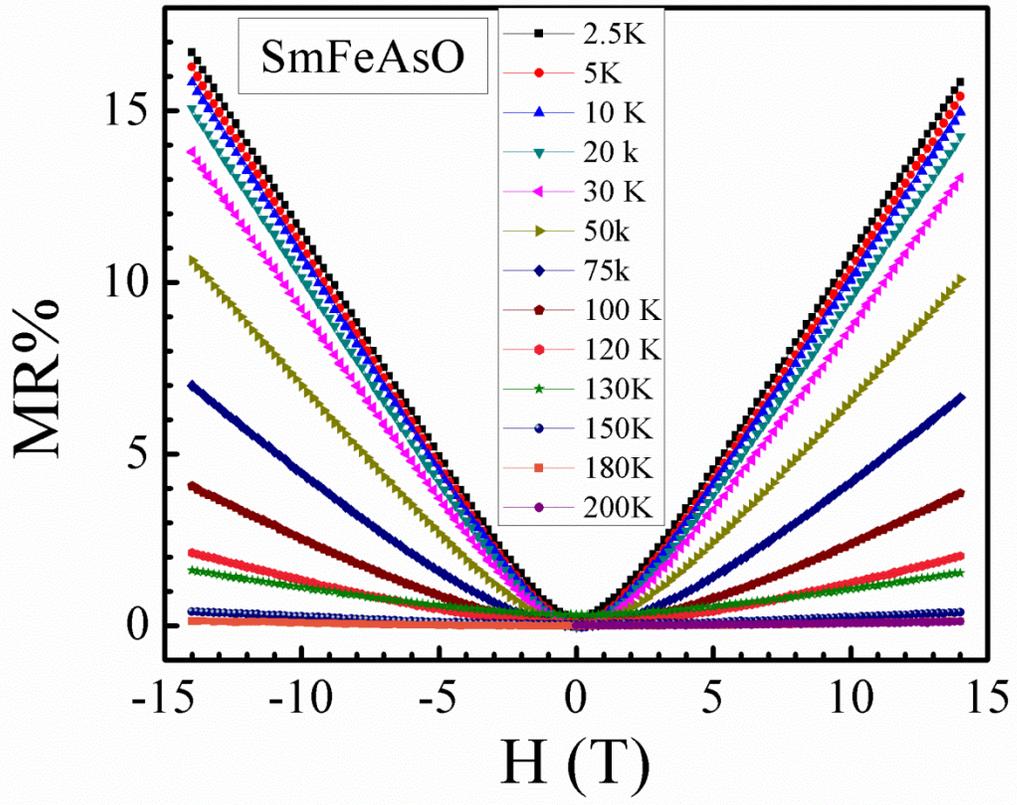

Fig. 2b

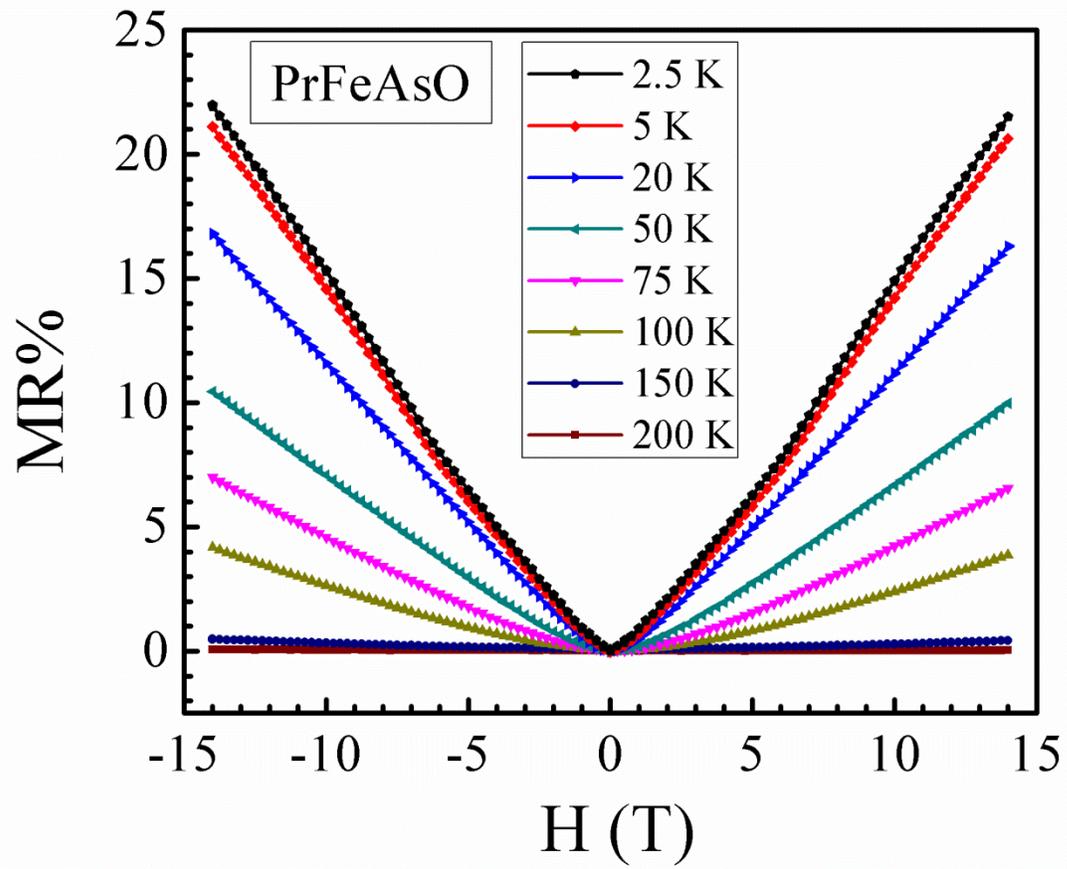

Fig. 3a

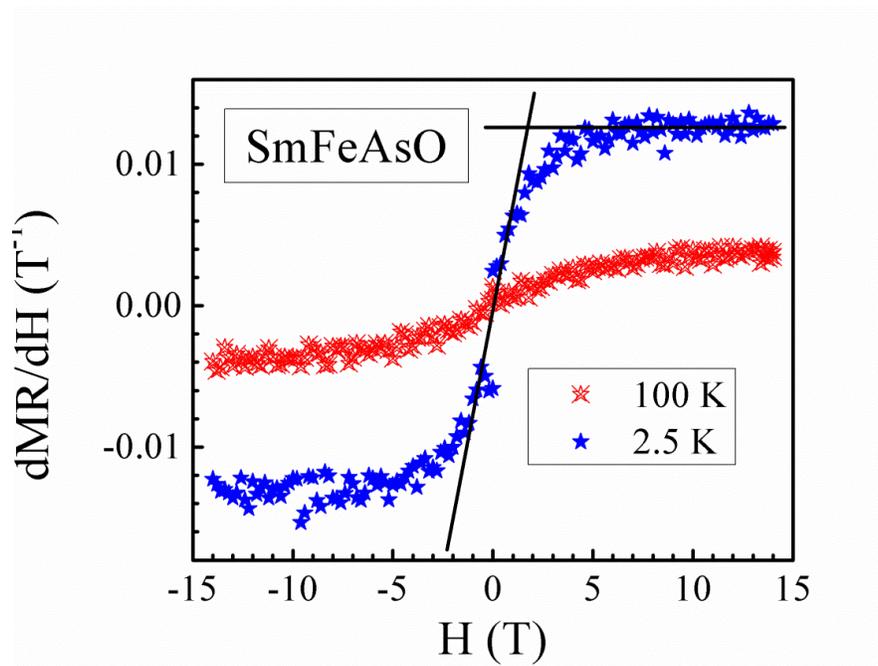

Fig. 3b

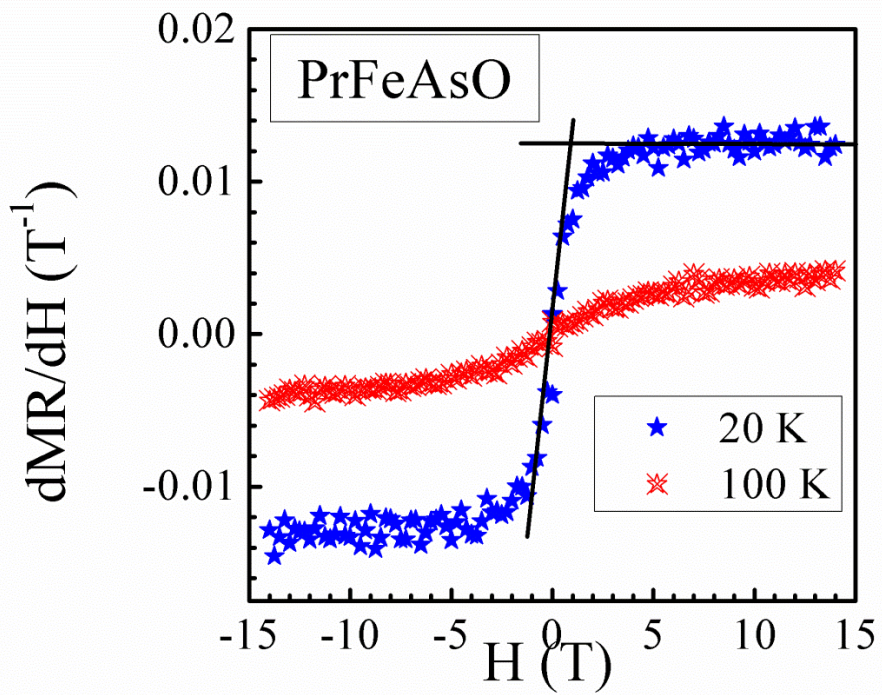

Fig. 3c

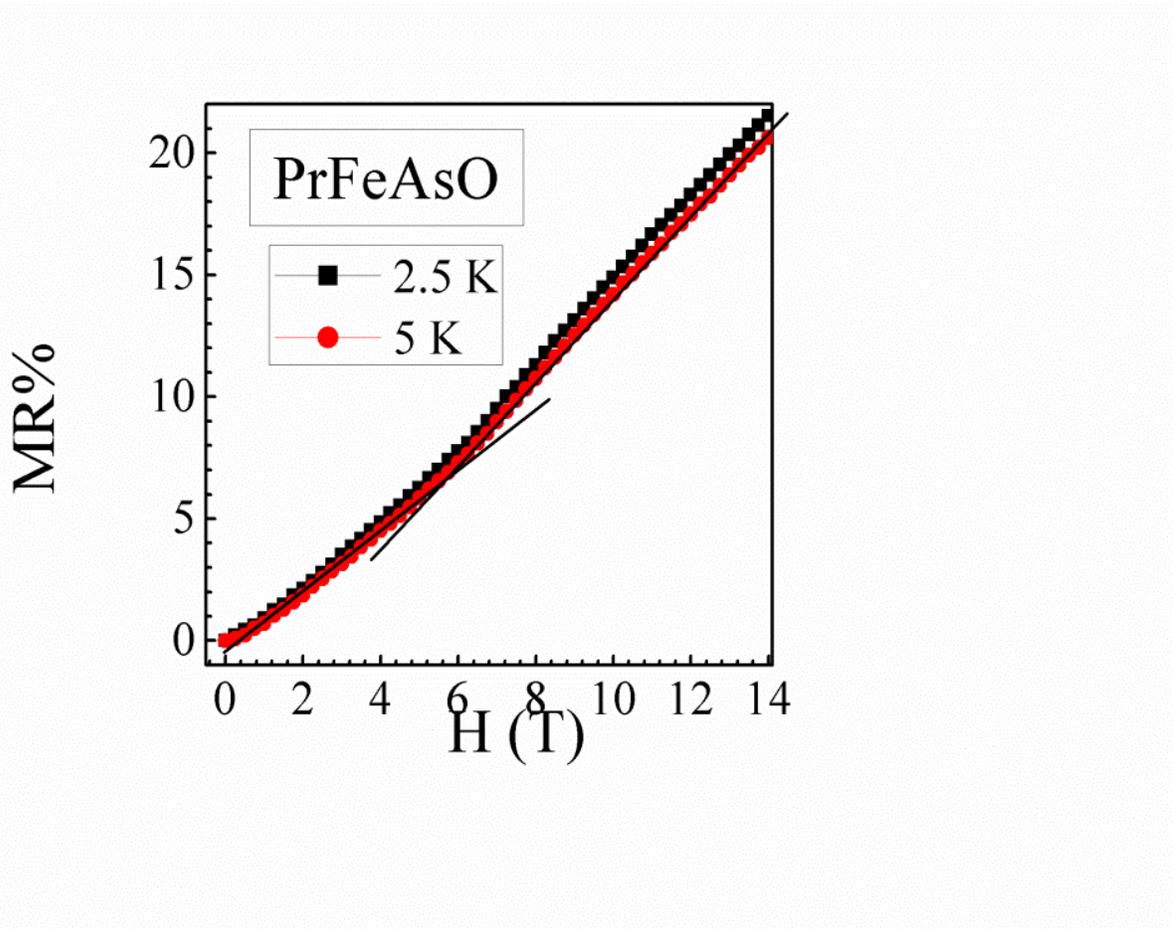